# Self-propelled film-boiling liquids

H. Linke [a,b,*], B. J. Alemán [a], L. D. Melling [a], M. J. Taormina [a], M. J. Francis [b], C.C. Dow-Hygelund [a], V. Narayanan [c], R. P. Taylor [a], and A. Stout [a]

[a]     *Materials Science Institute and Physics Department, University of Oregon, Eugene OR 97405-1274, U.S.A.*

[b]     *School of Physics, University of New South Wales, Sydney 2052, Australia*

[c]     *Department of Mechanical Engineering, Oregon State University, Corvallis, OR 97331, U.S.A.*

* Corresponding author: linke@uoregon.edu

           47.55.Dz   Drops and bubbles
           68.03.-g   Gas-liquid and vacuum-liquid interfaces

**Abstract**

We report that liquids perform self-propelled motion when they are placed in contact with hot surfaces with asymmetric (ratchet-like) topology. The pumping effect is observed when the liquid is in the film-boiling regime, for many liquids and over a wide temperature range. We propose that liquid motion is driven by a viscous force exerted by vapor flow between the solid and the liquid.

The ability to control the dynamics of liquids is crucial in applications such as lab-on-a-chip technology, ink-jet printing, thermal management, spray cooling and painting. It has previously been observed that millimeter-scale liquid droplets on substrates can move spontaneously due to an imbalance of surface tension forces (the Marangoni effect) [1-3], caused by a chemical [4-10], thermal [4, 11, 12], or electrical [13] gradient. In these systems, hysteresis forces [3] due to the wetting contact between droplet and surface limit the droplet speed to a few mm/s, unless additional power is supplied [14]. The resulting displacement is usually a few centimetres or less, because a finite gradient along the direction of motion is needed. Here we report self-propelled film-boiling droplets, which are separated from the supporting solid by a lubricating vapor layer. Droplet motion is driven by a thermal gradient perpendicular to the direction of motion, and is controlled by a saw tooth shaped substrate topology, without the need for a gradient along the flow direction (Movie 1). Millimeter-sized droplets accelerate at rates of up to $1 - 2$ m/s$^2$, can climb against inclines, and sustain speeds of order 5 cm/s over distances up to a meter. We propose that droplet motion is due to a viscous force exerted by vapor flow between the droplet and the asymmetric surface.

Figure 1(a) shows a sequence of video images demonstrating droplet motion perpendicular to the thermal gradient that powers the liquid motion. A droplet of refrigerant R134a (1,1,1,2 tetrafluorethane, boiling point $T_b$ = - 26.1 °C) was placed on a room-temperature, brass surface that was milled into a periodic, ratchet-like topology. The droplet is observed to move perpendicular to the ratchet features (in positive x-direction in Fig. 1(a)), reaching a terminal speed of several cm/s. The motion is sustained over the entire structured surface area, limited only by the evaporation time of the droplet (several tens of seconds). No external force or gradient (other than the vertical temperature difference between solid and liquid) is applied. We have observed this effect for all liquids we have investigated (including nitrogen, acetone, methanol, ethanol, water, and hexadecane, with boiling points ranging from − 196 °C to + 151 °C), provided the ratchet temperature is above the liquid's Leidenfrost point $T_L$ (about 20 °C for R134a on brass), where a vapor layer levitates the droplet. The effect is observed independent of the ratchet material and its precise geometry in the size ranges we have investigated (1 mm < $s$ < 3 mm, 0.1 mm < $d$ < 0.3 mm; see Fig. 1(a)).

Figure 2(a) shows the time evolution of the velocity of R134a droplets (radius $r \approx$ 1.5 mm, comparable to the liquid's capillary length $\kappa^{-1} = (\gamma/\rho g)^{1/2} = 1.1$ mm, where $\gamma =$ 15.5 mN/m is the liquid's surface tension at $T_b$, $\rho = 1377$ kg/m$^3$ is the mass density at $T_b$, and $g$ is the gravitational acceleration) on a horizontally levelled brass ratchet. Droplets were given an initial velocity $v_0$ before entering the ratchet (5 x 18 cm$^2$). The droplet trajectory was recorded using a video camera (frame rate 30/s), digitized using the software VideoPoint (Lenox, MA), and the velocity's x-component $v_x$ was calculated. Droplets that initially move in the negative x-direction in Fig. 1 ($v_{0x} < 0$) turn around and accelerate until they reach a terminal velocity $v_t$ (see Fig. 2(a) and Movies 3 and 4). The $v_x(t)$ data fit the equation

$$v_x(t) = \left(v_{0x} - \frac{a}{\beta/m}\right)e^{-\frac{\beta}{m}t} + \frac{a}{\beta/m} \quad (1)$$

which, for a droplet of mass $m$, is the solution of the equation of motion $m(dv_x/dt) = -\beta v_x(t) + F$, with a velocity-dependent drag force ($-\beta v_x$), a positive, constant accelerating force $F = ma$, and resulting terminal velocity $v_t = a/(\beta/m)$. $v_{0x}$, $a$ and $\beta/m$ are fit parameters.

In Figs. 2(b-d), we show results for droplets of R134a, ethanol ($T_b = 78.5$ °C, $\gamma =$ 18.2 mN/m, $\rho = 727$ kg/m$^3$, $\kappa^{-1} = 1.6$ mm), and water ($\gamma = 58.9$ mN/m, $\rho = 957$ kg/m$^3$, $\kappa^{-1}$ = 2.5 mm), respectively, as a function of the superheat $\Delta T = (T_R - T_b)$, where $T_R$ is the ratchet temperature (determined using NiCr/NiAl thermocouples inserted into horizontal, 2 mm wide holes approximately 4 mm underneath the ratchet surface). For each liquid, two temperature regimes can be distinguished. At temperatures within about 20 – 50 °C above the onset of self-propelled motion (Regime L), we find the highest accelerations (up to 1 –2 m/s$^2$), and the highest values for the drag parameter $\beta/m$. However, the results in Regime L vary from droplet to droplet, and the velocity of individual droplets can fluctuate significantly (see inset to Fig. 2(a)). At higher temperatures (Regime H), we observe significantly lower $a$ and $\beta/m$, and much smaller fluctuations.

We speculate that in Regime L droplets on a ratchet are not fully in the film-boiling regime, and nucleate boiling events introduce fluctuations. This interpretation is supported by our observation that in Regime L the ratchet-cleaning method influences the droplet dynamics (Fig. 2 (b-c)), and by the fact that a liquid's Leidenfrost point is known to vary substantially with surface roughness and contamination [15, 16]. In Regime H, however, we find that surface contamination has little influence on $a$ and $\beta/m$, consistent with the liquid and substrate being fully separated by a vapor layer.

The orientation of the ratchet plane relative to gravity is not critical: droplets or slugs placed in a flat-bottomed, open channel (width 2-4 mm) with ratcheted, vertical side walls, accelerate against significant inclines, with the effect becoming stronger with decreasing channel width (Movie 5). This observation suggests that the role of gravity is limited to keeping the liquid in contact with the ratchet.

For ratchet periods $s \approx 1 - 2$ mm we do not observe motion for droplets with radius $r < 0.3\ s$. Droplets spanning multiple $s$ do accelerate (Movies 5 and 6), but puddles tend to break up into smaller droplets once $r$ exceeds several $\kappa^{-1}$. If confined into a channel we observe acceleration of slugs tens of millimetres in length, interacting with many ratchet periods simultaneously (Movie 5).

To explain our observations, we propose that an asymmetry in the vapor-flow pattern underneath the droplet, induced by the substrate topology, exerts a viscous force on the droplet. In the following we present a model based on this notion.

A droplet placed on a ratchet (see Fig. 3(a)) tends to curve concavely around the tops of the ridges (point A) while assuming a convex shape elsewhere. This variation in droplet curvature implies a pressure differential along the vapor layer as explained in the following. The local difference between the droplet's internal pressure $p_i$ (assumed constant along the bottom surface) and the pressure in the vapor film is given approximately by the Laplace pressure $\Delta p = \gamma/R$, where $R$ is the local radius of the curvature (assuming no curvature parallel to the ratchet ridges) [3]. A concave surface shape (near point A) corresponds to a curvature $R_A < 0$ and $p_A > p_i$, while the convex curvature at points $B_1$ and $B_2$ implies $R_B > 0$ and $p_B < p_i$, such that $p_A > p_B$. We therefore expect net vapor flow from point A to points $B_1$ and $B_2$. Flow from A to $B_2$ is expected to create a viscous force in forward direction, which we estimate below. In contrast, vapor flowing from A "backward" can escape sideways along the wide ratchet grooves (into and out of the page in Fig. 3(a)), because of the small flow resistance in this direction [17]. Therefore, net forces due to vapor flow between A and $B_1$ should be relatively small.

The force exerted by the vapor on the liquid between points A and $B_2$ has two components. First, a forward shear force due to Poiseuille vapor flow caused by the pressure differential $\Delta P = (p_A - p_B)$. Using nonslip boundary conditions and a parallel-plate model, the horizontal component of this force is [18]

$$F = 0.5\, A_{eff}\, h\, |dP/dx|\, \cos\theta \tag{2}$$

where $A_{eff}$ is the total area over which this force contributes (depending on droplet size, multiple ratchet periods are involved), $h$ is the thickness of the vapor layer in this area, and $\theta$ is defined in Fig. 3(b). Second, if the droplet glides with $v_x$ relative to the substrate there is a viscous drag force given by [18]

$$-\beta v_x = -(\eta\, A_{eff}/h)\, v_x \tag{3}$$

where $\eta$ is the vapor's viscosity.

In Fig. 4(a) we show measured values of $a$ for water droplets as a function of their volume $V$. For droplets that are too small to cover about three full ratchet periods ($r < 1.5\, s$, where $s = 1.5$ mm), there is considerable scatter in the data. We therefore focus on the range $V > 50$ μl ($r > 2.3$ mm) when we compare data and model in the following. To calculate $a$ from Eq. (2), we estimate the fraction $\alpha = A_{eff}/A_c$, where $A_c(V) = (5/4)\, \kappa^{1/2}\, V^{5/6}$ is the droplet's volume-dependent contact area [19], by measuring the distance $l$ between points of type A and $B_2$ in high-resolution photographs of droplets with $r \approx 1.5\, s$ and find $\alpha \approx 0.6$ for water. The value of $h$ is typically $10 - 100$ μm on flat surfaces [20, 21], but varies with position on a ratchet surface (see Fig. 3(a)). We determine $h$ from the measured values for $\beta/m$ using Eq. (3), neglecting contributions to the drag force from areas outside $A_{eff} = \alpha A_c(V)$. We find that $h$ varies weakly with droplet size for $V > 50$ μl (see Fig. 4 (b)) and use the averaged value in this range, $h = 10.2$ μm, for the further analysis. We determine $R_A$ and $R_B$ from high-resolution photographs to find pressure gradients $dP/dx \approx (p_A - p_B)/l$ of order $10^2$ Pa/mm [22], confirming that the hydrostatic variations in $p_i$ of order of $\rho g d/s \approx 3$ Pa/mm are negligible.

The droplet acceleration $a = F/m$ of water droplets as a function of $V$ derived from Eq. (2) is plotted in Fig. 4(a), along with the observed data. The model quantitatively agrees with the data within the uncertainty of the calculation, which is mainly due to error in α, $R_A$, $R_B$, and $h$. Note that the expression for $A_c(V)$ used here [19], as well as expressions proposed more recently [21], assume circular droplets, while our droplets are typically elongated in the *x*-direction. This may account for why the model deviates from the data for larger droplets ($r > 3$ mm) which are more likely to elongate. Consistent with experiment, the model predicts that droplets with $r < 0.5\ s$, for which no curvature differentials are expected, should not accelerate. For droplets with $s \approx r$, the droplet curvature changes with time in a complex manner, which may explain the scatter in the data for *a* in this range.

The above model is relevant to Regime H. We propose that in Regime L nucleate boiling occurs near the ratchet ridges, explaining the occasionally observed high values of $\beta/m$ (hysteresis forces during wetting events introduce additional drag), and of *a* (due to the additional bursts of vapor flow). Nucleate boiling may also contribute to the acceleration of a liquid nitrogen droplet across the edge of a piece of tape (VWR scientific tape) placed on a brass surface (see Fig. 1(b) and Movie 8). We observe no acceleration when we use a brass step of equal height, and propose that vapor flow associated with a nucleate boiling event at the tape edge (which is expected to cool much faster than brass, promoting nucleate boiling) exerts a forward force on the droplet (see Fig. 3(c)).

Thermocapillary flow [3] along the droplet's bottom surface may also be important. For instance, during a nucleate boiling event at a tape edge (Fig. 1(b)), the liquid surface closest to the wetting point is likely to be heated above $T_b$, inducing thermocapillary flow along the x-direction, away from the wetting point (Fig. 3(c)). Due to the broken symmetry at the step this may result in net pumping action, with the hysteresis forces at the wetting point providing a reaction force. A similar mechanism may be at work in periodic ratchets in Regime L. In Regime H, where frictional forces between liquid and solid are almost absent, a mechanism for direct momentum exchange between solid and liquid is less obvious, but may occur when the ratchet topology forces a change in the direction of thermocapillary flow (at point B in Fig. 3(b)). We speculate that by relying on thermocapillary flow it may be possible to use a ratchet-like topology to propel droplets at $T_R < T_L$ if a ratchet with superhydrophobic coating is used to mitigate friction.

The ratchet effect [23] reported here may be used to construct pumps consisting of channels with ratchet-shaped inner walls. Such ratchet pumps could be powered by waste heat, making them attractive for use in millimeter-scale closed loop, two-phase cooling systems with no moving parts and no external power need, for example for microprocessor cooling. It remains to be explored whether the pumping is sustained for ratchet dimensions below the millimeter-scale, enabling microfluidic applications.

**Acknowledgments:** Technical help by Jack Sandall, Ralf Müller, Gwynne Engelking and Leif Karlström, and useful discussions with Russell Donnelly, Brian Long, Sandra Troian, Julia Yeomans, and Kevin Young. Financially supported by NSF CAREER, the Australian Research Council, the McNair Foundation (B.J.A.), NSF-REU and Mentor Graphics (L.M., A.S.).

Supporting movies are available at http://darkwing.uoregon.edu/~linke/supporting .

**Figure legends**

**Fig. 1.** (a) Video-sequence (time interval 32 ms) of a droplet of R134a on a horizontally levelled, brass surface with ratchet-like topology ($d = 0.3$ mm, $s = 1.5$ mm). See Movie 2. (b) Droplet of liquid nitrogen on a flat brass surface moving with a small initial velocity towards a piece of tape (shaded), accelerating on interaction with the tape edge (see Movie 8).

**Fig. 2 (Color online).** (a) Velocity evolution of a droplet of R134a ($r \approx 1.5$ mm) on the brass ratchet shown in Fig. 1, held at $T_R = 70$ °C (main figure) and $T_R = 22$ °C (inset. The average displacement over three data points (33 ms apart) was used to determine instantaneous velocity. Solid lines are fits to Eq. (1). (b), (c), and (d) show the fit parameter $a$ for R134a, ethanol, and water, respectively, as a function of $\Delta T = (T_R - T_b)$. Crosses: ratchet surface was cleaned manually using Kimwipes with brass polish (Wright, Keane, New Hampshire) and rinsed with de-ionized water. Open circles: data taken after additional sonication (several minutes each in acetone, isopropyl alcohol and methanol, then rinsed in de-ionized water). For each liquid we distinguish a low-temperature regime (L) and a high-temperature regime (H).

**Fig. 3 (Color online).** (a) Photograph of the liquid-vapor-solid interface (side view) of a droplet of film-boiling water on a brass ratchet ($T_R = 460$ °C), showing concave curvature near point A, and convex curvature near $B_1$, $B_2$. The scale bar is 1.5 mm. (b) The vapor flow (black arrows) generated by the pressure differential $\Delta p = p_A - p_B$ in a vapor layer of thickness $h$ is expected to generate a forward force onto the droplet. Vapor flowing from A backwards is expected to escape sideways along the ratchet grooves (see Supplementary Video 6). Red arrows indicate expected thermocapillary flow. (c) A nucleate boiling event near the tape edge in Fig. 1(b) is expected to lead to vapor flow (black arrows) and to thermocapillary flow (red arrows).

**Fig. 4 (Color online).** (a) Acceleration $a$ as a function of droplet volume, controlled with a pipette ($T_R = 460\ °C$). The full line is the model prediction for $a = F/m$ based on Eq. (2) for droplets large enough to cover at least three ratchet periods ($r > 1.5\ s$) as indicated by the cartoon. **(b)** Values for $h$ determined from measured $\beta/m$ (Eq. 3), using the vapor viscosity $\eta$ near the mean of $T_R$ and $T_b$ ($\eta = 19.4\ \mu Pa\ s$). The horizontal line indicates the volume-averaged $h$ used in the model shown in (a).

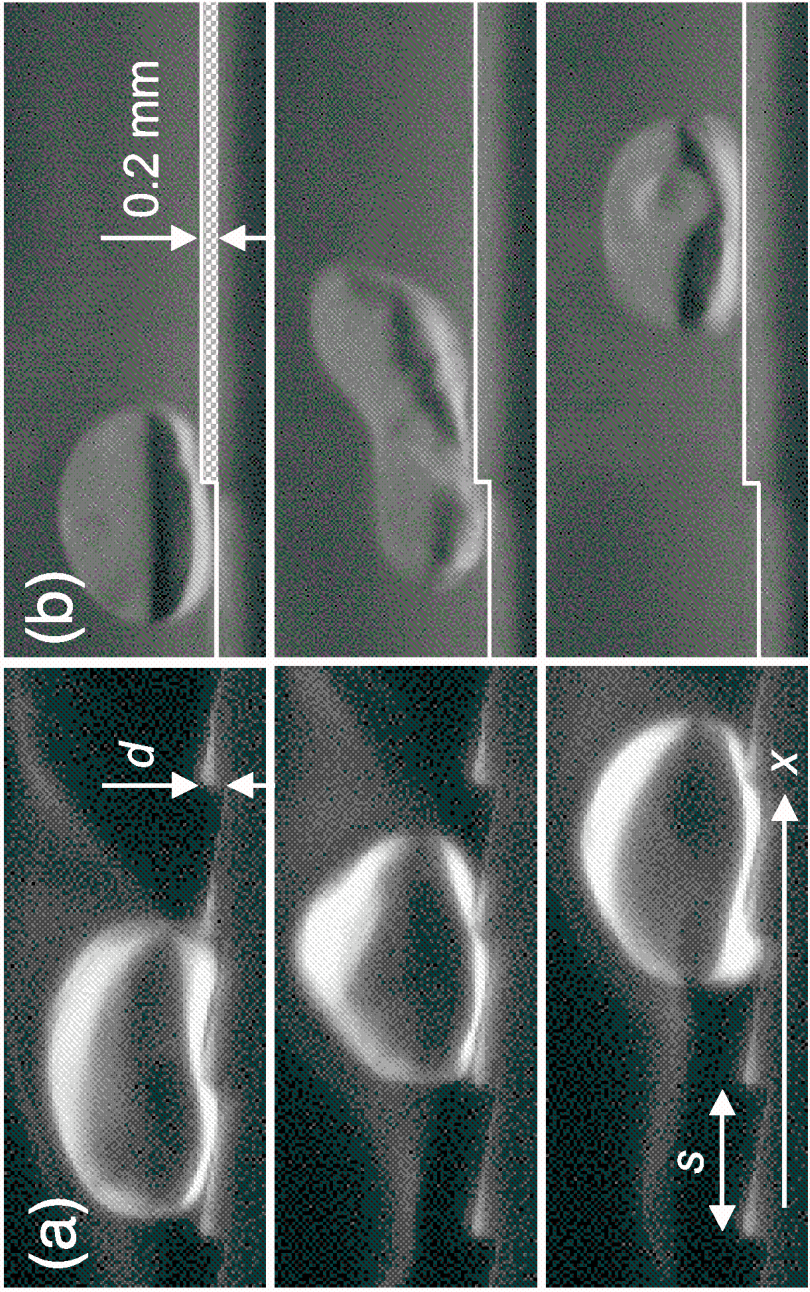

Fig. 1
Linke et al.
Physical Review Letters

Figure 2
Linke et al.
Physical Review Letters

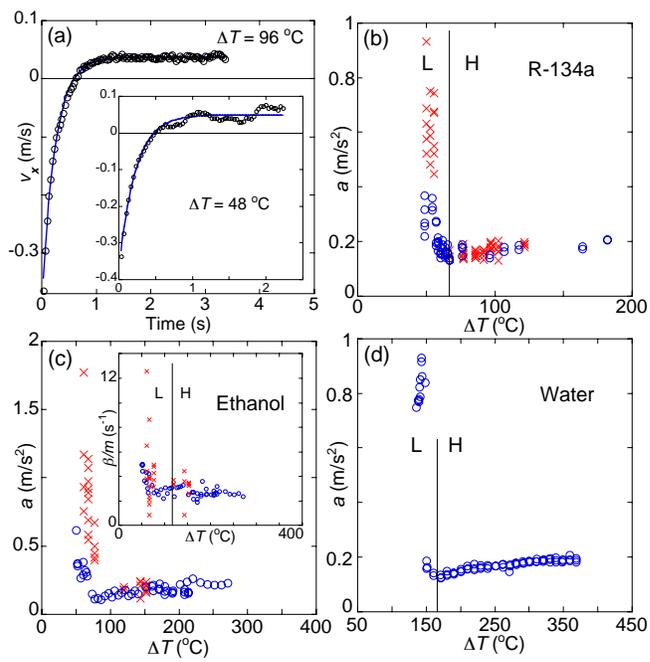

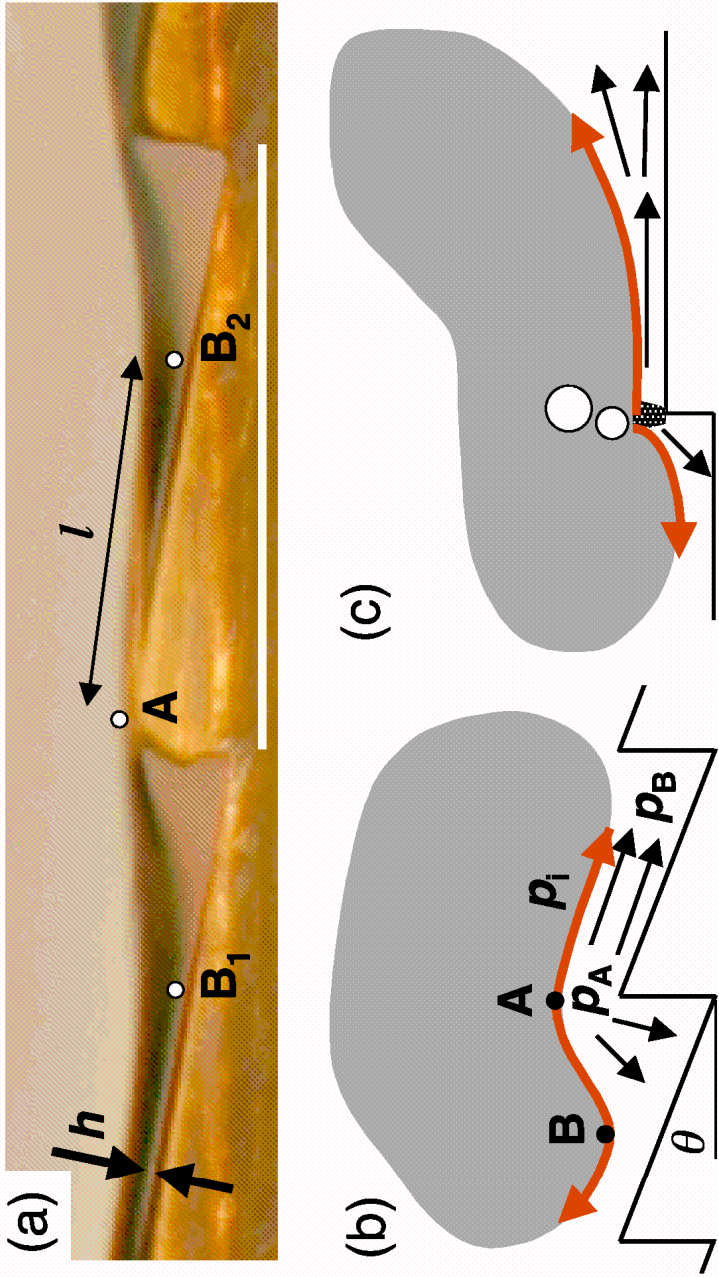

Fig. 3
Linke et al.
Physical Review Letters

Figure 4
Linke et al.

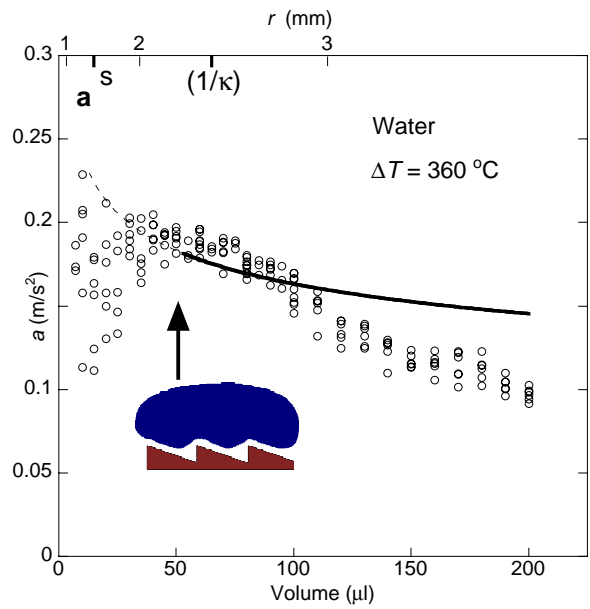

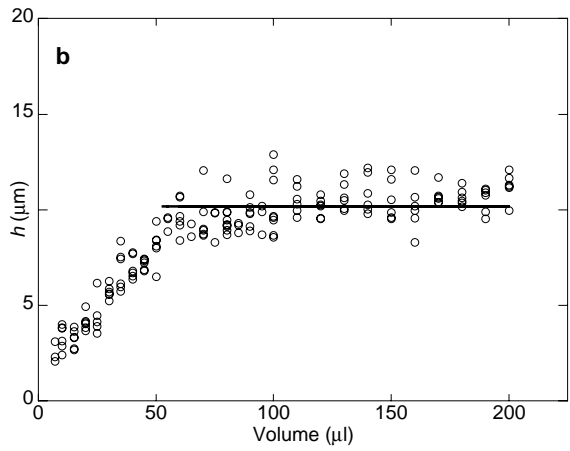